\documentclass[aps,showpacs,prl,superscriptaddress,]{revtex4}
\usepackage{amsmath}
\usepackage{amsfonts}
\usepackage{amssymb}
\usepackage{bm}
\usepackage{graphicx}

\begin{document}

\title{Breather and interacting soliton and periodic waves for modified
KdV equation}
\author{Vladimir I. Kruglov}
\affiliation{Centre for Engineering Quantum Systems, School of Mathematics and Physics,
The University of Queensland, Brisbane, Queensland 4072, Australia}
\author{ Houria Triki}
\affiliation{Radiation Physics Laboratory, Department of Physics, Faculty of Sciences,
Badji Mokhtar University, P. O. Box 12, 23000 Annaba, Algeria}

\begin{abstract}
We present the discovery of a class of exact spatially localized as well as
periodic wave solutions within the framework of the modified Korteweg-de Vries
equation. This class comprises breather and interacting soliton
solutions as well as interacting periodic wave solutions. The
functional forms of these solutions include a joint parameter which can take
both positive and negative values of unity. It is found that the existence
of those closed form solutions depend strongly on whether the cubic
nonlinearity parameter should be considered positive or negative. The
derived wave structures show interesting properties that may find 
practical applications.
\end{abstract}

\maketitle
\affiliation{$^{1}${\small Radiation Physics Laboratory, Department of Physics, Faculty
of Sciences, Badji Mokhtar University, P. O. Box 12, 23000 Annaba, Algeria}\\
$^{2}${\small Centre for Engineering Quantum Systems, School of Mathematics
and Physics, The University of Queensland, Brisbane, Queensland 4072,
Australia}}

The modified Korteweg-de Vries (mKdV) equation is an important model which
applies to the description of wave dynamics in different physical systems,
such as soliton propagation in lattices \cite{R1}, meandering ocean currents 
\cite{R2}, the dynamics of traffic flow \cite{R3,R4,R5}, nonlinear Alfv\'{e}%
n waves propagating in plasma \cite{R6}, and ion acoustic soliton
experiments in plasmas \cite{R7}. This model is also relevant for nonlinear
waves in distributed Schottky barrier diode transmission lines \cite{R8} and
internal waves in stratified fluids \cite{R9}. Recent results have also
demonstrated that the propagation of optical pulses consisting of a few
cycles in Kerr-type media can be described beyond the slowly varying
envelope approximation by using the mKdV equation \cite{R10,R11,R12,R13}.
Additionally, this equation has gained further importance recently, mainly
because of its effectiveness in modeling supercontinuum generation in
nonlinear optical fibers \cite{R14}.

The mKdV equation has been successfully used to model the evolution of long
waves in the critical case of vanishing quadratic nonlinearity \cite{Pel}.
Due to the wide-ranging potential applications of this nonlinear wave
evolution equation, various powerful methods have been employed to search
for its explicit solutions. This because wave solutions are helpful for a
better understanding of physical phenomena modeled by this equation such as
the stability of nonlinear wave propagation. Having solutions in analytic
form is relevant not only for determining certain important physical
quantities and serving as diagnostics for simulations but also even for
comparing experimental results with theory. Particularly, Wadati derived the
exact N-soliton solutions of the mKdV equation by using the
inverse-scattering transform scheme \cite{Wadati}. In addition, Kevrekidis 
\textit{et al}. \cite{Kevrekidis} obtained some classes of periodic
solutions of this model. Another class of nonlinear wave solutions that
conserve their energy during evolution -breathers (oscillatory wave
packets)- has been also found for this model (see, e.g., \cite{Lamb,G,An}).
The focusing and defocusing mKdV equations with nonzero boundary conditions
are studied for inverse scattering transforms with matrix Riemann-Hilbert
problem in Ref. \cite{Zhang}.

In this Letter, we predict the existence of three types of nonlinear wave
structures through discovery of physically important exact solutions of the
mKdV model that includes the cubic nonlinearity term with either positive or
negative sign. Significant classes of breather and interacting  soliton pairs 
and periodic waves are presented for the first time. Remarkably, the breather and 
interacting periodic waves formation are observed in the case of a negative
coefficient of the cubic nonlinear term while the pair of interacting
soliton solution is formed when this coefficient is positive.

We start by considering the mKdV equation in standard dimensionless form as%
\begin{equation}
u_{t}+u_{xxx}+6\mu u^{2}u_{x}=0,  \label{1}
\end{equation}%
where $u(x,t)$ is the real function. The parameter $\mu=\pm 1$ denotes
the type of nonlinearity, i.e., $+1$ for focusing type of nonlinearity and $%
-1$ for defocusing nonlinearity. The mKdV is a fully integrable equation
which means that it has an infinite number of conserved invariants \cite%
{Miur}.

Two soliton branches exist for the mKdV equation (\ref{1}) in the case of a
positive coefficient of the cubic nonlinear term (i.e., $\mu =1$)$,$ they
are defined as 
\begin{equation}
u(x,t)=a+\frac{b^{2}}{\Lambda \cosh (b(x-x_{0}-vt)+\psi_{0})+2a},  \label{2}
\end{equation}%
where $\Lambda=\pm \sqrt{4a^{2}+b^{2}}$, $v=6a^{2}+b^{2}$, and $a$, $b$, $%
x_{0}$ are the arbitrary constants. In the limit when $a\rightarrow 0$ the
solution in Eq. (\ref{2}) tends to the familiar \textrm{\ sech}-shaped
soliton family which is known to be stable with respect to small
perturbations (see, e.g., \cite{Pego}). In the case with $\mu=1$ also exists
the algebraic solitary wave solution as 
\begin{equation}
u(x,t)=p-\frac{p}{p^{2}(x-x_{0}-vt)^{2}+1/4},  \label{3}
\end{equation}
where $v=6p^{2}$, and $p$, $x_{0}$ are the arbitrary constants. We note that
the N-soliton solution of (1) can be obtained using the Darboux transform: 
\begin{equation}
u(x,t)=-i\frac{\partial}{\partial x}\ln\frac{W(\Psi_{1x},\Psi_{2x},...,%
\Psi_{Nx})} {W(\Psi_{1},\Psi_{2},...,\Psi_{N})},  \label{4}
\end{equation}
where $W$ is the Wronskian for $N$ eigenfunctions $\Psi_{j}$ in the
denominator and for their spatial derivatives $\Psi_{jx}$ in the numerator.
This formal expression has inner symmetries and the properties of determinants
which can be used for construction corresponding solutions \cite{Mat}.

The breather's expression for $\mu=1$ was obtained from inverse scattering
transform by Pelinovsky and Grimshaw in \cite{Pel} and also in Ref. \cite%
{Slu}. Note that the breather has much more complicated dynamics than
soliton. Recently Slunyaev and Pelinovsky \cite{Slunyaev} also presented an
explicit breather solution for the mKdV equation (\ref{1}) with $\mu =1$\ as 
\begin{equation}
u(x,t)=2pq\left( \frac{p\sinh (\theta )\sin (\phi )-q\cosh (\theta )\cos
(\phi )}{p^{2}\sin ^{2}(\phi )+sq^{2}\cosh ^{2}(\theta )}\right) ,  \label{5}
\end{equation}%
when $s=-1.$ Here $\theta $ and $\phi $ which control the wave envelope and
the inner wave respectively, are given by 
\begin{equation}
\theta =p(x-x_{0})+p(3q^{2}-p^{2})t+\theta _{0},  \label{6}
\end{equation}%
\begin{equation}
\phi =q(x-x_{0})+q(q^{2}-3p^{2})t+\phi _{0},  \label{7}
\end{equation}
where $p$, $q$ and $\theta _{0}$, $\phi _{0}$, $x_{0}$ are the arbitrary
real parameters. We have remarked that this breather solution still also
existing for the case when $s=+1$ with the same relations of $\theta \ $and $%
\phi $ as those given in Eqs. (\ref{6}) and (\ref{7}). Thus the mKdV
equation (\ref{1}) with the focusing type of nonlinearity possesses two
exact breather solutions (\ref{5}) corresponding to the values $s=\pm 1$.

\begin{figure}[h]
\includegraphics[width=1\textwidth]{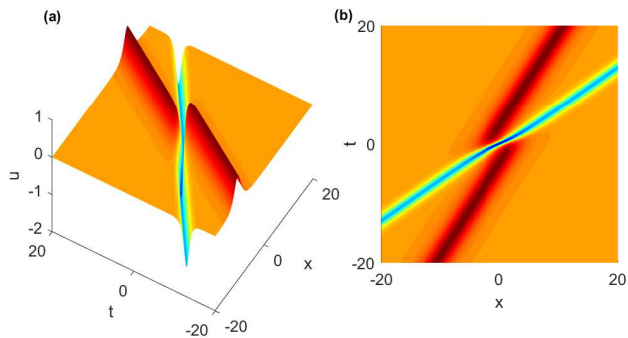}
\caption{(a) Evolution of the interacting soliton solution of mKdV
	equation defined by Eq. (\ref{8}) ( $\mu $ $=1$) for the values $p=1,$\ $q=0.2$\
	and $s=1$. (b) The corresponding contour plot of the interacting soliton
	solution in (a).}
\label{FIG.1.}
\end{figure}

In what follows, we report what is to our knowledge the analytical
demonstration of a new class of breather, interacting 
periodic and interacting soliton solutions for the mKdV equation
with either focusing or defocusing types of nonlinearity. More precisely,
using the inverse scattering method (the details can be published somewhere)
we have found three types of exact solutions for the modified KdV equation (%
\ref{1}) with $\mu =\pm 1.$ These solutions are given in Eq. (\ref{8}) (for $%
\mu =1$), and Eqs. (\ref{13}) and (\ref{18}) (for $\mu =-1$). A first class
of localized waves in the form of a pair interacting solitons is
obtained for Eq. (\ref{1}) in the case of focusing cubic nonlinearity (i.e., 
$\mu =1$) as 
\begin{equation}
u(x,t)=2pq\left( \frac{p\sinh (\theta )\sinh (\phi )-q\cosh (\theta )\cosh
(\phi )}{p^{2}\sinh ^{2}(\phi )+sq^{2}\cosh ^{2}(\theta )}\right) ,
\label{8}
\end{equation}%
where $s=\pm 1$ and $\theta $, $\phi $ are given by 
\begin{equation}
\theta =p(x-x_{0})-p(3q^{2}+p^{2})t+\theta _{0},  \label{9}
\end{equation}%
\begin{equation}
\phi =q(x-x_{0})-q(3p^{2}+q^{2})t+\phi _{0}.  \label{10}
\end{equation}

Here and below $p$, $q$ and $\theta _{0}$, $\phi _{0}$, $x_{0}$ are the
arbitrary real parameters. The wave solution (\ref{8}) represents a
pair of interacting bipolar solitons whose dynamic features are delineated
in Fig. 1 (with $\mu =1$) for the case $p=1,$ $q=0.2$
and $s=1$. Here we choose the initial soliton position $x_{0}$ and phases $\theta _{0}$ and $\phi _{0}$ equal to zeros. Figure 1 shows the overtaking interaction between two solitons of opposite polarity for $u(x,t)$ given in Eq. (8), from
which we can see that the solitonic amplitudes and shapes have not changed
after the interaction. Compared with the well-known 
sech-shaped solitary wave which is a single soliton structure,
this novel mKdV soliton solution takes the form of a pair of interacting
solitons, which can interact purely elastically and behave like independent
solitons far from the meeting point. 
We notice that the wave solution (\ref{8}) with $s=-1$ also takes the shape of a pair of interacting solitons with opposite polarity.

In the case $q=-p$ and $s=1$ the solution given in Eq. (\ref{8}) has the form of modified soliton solution traveling with dimensionless velocity $v$. This modified soliton solution is given by 
\begin{equation}
u(x,t)=\frac{4q\cosh (\theta _{0}+\phi _{0})}{\cosh (2q\xi +2\phi_{0})
+\cosh (2q\xi -2\theta _{0})},  \label{11}
\end{equation}%
where $\xi =(x-x_{0})-vt$ and $v=4q^{2}$. 
\begin{figure}[h]
\includegraphics[width=1\textwidth]{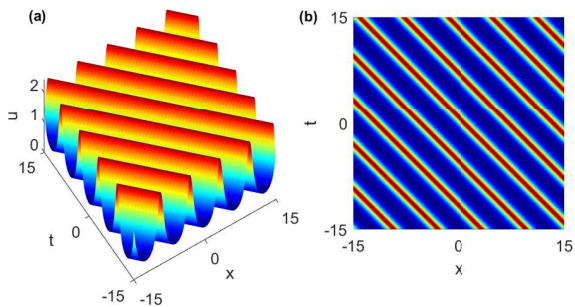}
\caption{ (a) Evolution of periodic traveling wave solution of mKdV
equation defined by Eq. (\ref{16}) ( $\mu $ $=-1$) for the values $p=0.5$,
$\phi_{0}=0$ and $\theta _{0}=\pi /4$. (b) The corresponding contour plot of the periodic wave solution in (a).}
\label{FIG.2.}
\end{figure}
In the particular case $\theta _{0}+\phi _{0}=0$ the solution (\ref{11})
reduces to soliton solution in (\ref{2}) with $a=0$, $b=2q$ and $\psi
_{0}=2\phi _{0}$.

In the case $q=p$ and $s=1$ the solution given in Eq. (\ref{8}) also has the form of modified soliton solution traveling with dimensionless velocity $v$. This modified soliton solution is given by 
\begin{equation}
u(x,t)=\frac{-4q\cosh (\theta _{0}-\phi _{0})}{\cosh (2q\xi +2\phi_{0})
+\cosh (2q\xi +2\theta _{0})},  \label{12}
\end{equation}%
where $\xi =(x-x_{0})-vt$ and $v=4q^{2}$. 
In the particular case $\theta _{0}-\phi _{0}=0$ the solution (\ref{12})
reduces to soliton solution in (\ref{2}) with $a=0$, $b=2q$ and $\psi
_{0}=2\phi _{0}$.

We have obtained another exact solution for the mKdV equation (\ref{1}) with 
$\mu =-1$ in the form of an interacting periodic wave solution as 
\begin{equation}
u(x,t)=2pq\left( \frac{p\sin (\theta )\sin (\phi )+q\cos (\theta )\cos (\phi
)}{p^{2}\sin ^{2}(\phi )-sq^{2}\cos ^{2}(\theta )}\right) ,  \label{13}
\end{equation}%
where $s=\pm 1$ and $\theta $, $\phi $ are 
\begin{equation}
\theta =p(x-x_{0})+p(3q^{2}+p^{2})t+\theta _{0},  \label{14}
\end{equation}%
\begin{equation}
\phi =q(x-x_{0})+q(3p^{2}+q^{2})t+\phi _{0}.  \label{15}
\end{equation}

In the case $q=-p$ and $s=-1$ the solution given by Eq. (\ref{13}) takes the
shape of periodic traveling  wave, 
\begin{equation}
u(x,t)=\frac{2p\cos (\theta _{0}+\phi _{0})}{\sin ^{2}(p\xi -\phi _{0})+\cos
^{2}(p\xi +\theta _{0})},  \label{16}
\end{equation}%
where $\xi =(x-x_{0})-vt$ and $v=-4p^{2}$. The denominator in (\ref{16}) is
zero only in the case when $\theta _{0}+\phi _{0}=\pi /2+\pi n$ with $%
n=0,\pm 1,\pm 2,...$~. If this denominator is zero the numerator in (\ref{16}%
) is zero as well. The solution given in (\ref{16}) is not singular
when $\phi _{0}+\theta _{0}\neq \pi /2+\pi n$ with $n=0,\pm 1,\pm 2,...$~.
In the particular case $\theta _{0}+\phi _{0}=0$ this periodic wave solution reduces
to constant solution $u(x,t)=2p$. 

In the case $q=p$ and $s=-1$ the solution given by (\ref{13}) also takes the
shape of periodic traveling wave,     
\begin{equation}
u(x,t)=\frac{2p\cos (\theta _{0}-\phi _{0})}{\sin ^{2}(p\xi +\phi _{0})+\cos
^{2}(p\xi +\theta _{0})},  \label{17}
\end{equation}%
where $\xi =(x-x_{0})-vt$ and $v=-4p^{2}$. The denominator in (\ref{17}) is
zero only in the case when $\theta _{0}-\phi _{0}=\pi /2+\pi n$ with $%
n=0,\pm 1,\pm 2,...$~. If this denominator is zero the numerator in (\ref{17}) is zero as well. The solution given in (\ref{17}) is not singular
when $\theta _{0}-\phi _{0} \neq \pi /2+\pi n$ with $n=0,\pm 1,\pm 2,...$~.
In the particular case $\theta _{0}-\phi _{0}=0$ this periodic wave solution reduces
to constant solution $u(x,t)=2p$.
\begin{figure}[h]
\includegraphics[width=0.6\textwidth]{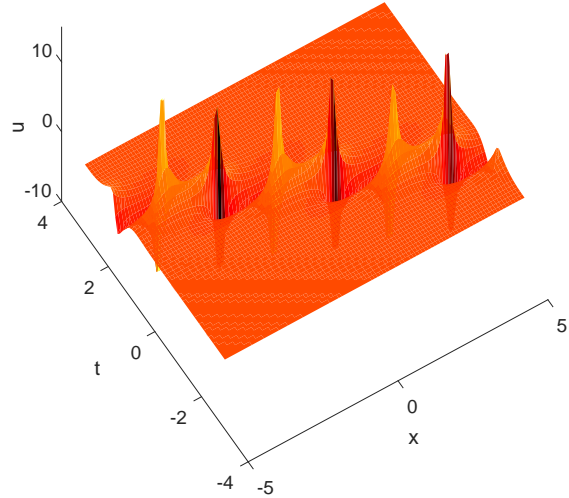}
\caption{Breather solution of mKdV equation defined by Eq. (\ref{18}) 
($\mu=-1$) for the values $p=1,$ $q=0.98$ and $s=-1$.}
\label{FIG.3.}
\end{figure}
Figure 2 depicts an example of the nonlinear wave solution (\ref{16}) 
(with $\mu =-1$) for the case $p=0.5$, $\phi_{0}=0$ and $\theta
_{0}=\pi /4$. Here the initial parameter $x_{0}$ is chosen equal to zeros. We observe that the nonlinear waveform (\ref{16}) presents an oscillating behaviour during the process of wave evolution.

We have also found that the mKdV equation (\ref{1}) with the defocusing type
of nonlinearity (i.e., $\mu =-1$) has an exact breather solution of
the form, 
\begin{equation}
u(x,t)=2pq\left( \frac{p\sin (\theta )\sinh (\phi )+q\cos (\theta )\cosh
(\phi )}{p^{2}\sinh ^{2}(\phi )-sq^{2}\cos ^{2}(\theta )}\right) ,
\label{18}
\end{equation}%
where $s=\pm 1$ and $\theta $, $\phi $ are 
\begin{equation}
\theta =p(x-x_{0})+p(p^{2}-3q^{2})t+\theta _{0},  \label{19}
\end{equation}%
\begin{equation}
\phi =q(x-x_{0})+q(3p^{2}-q^{2})t+\phi _{0}.  \label{20}
\end{equation}

To our knowledge, exact solutions (\ref{8}), (\ref{13}) and (\ref{18}) to the
mKdV equation with the focusing and defocusing type of nonlinearity are
firstly reported in this paper. From these important results, one may
conclude that a physical system described by the mKdV equation could allow a
breather evolution in either the focusing or the defocusing nonlinearity. A
typical example of the singular breather profile for solution in Eq. (\ref{18}) is shown Fig. 3 (with $\mu =-1$) for
the values $p=1,$ $q=0.98$ and $s=-1$. For this case, the initial parameters 
$x_{0}$, $\theta _{0}$ and $\phi _{0}$ are chosen equal to zeros. It can be
seen that the breather structure has a nontrivial periodical behaviour while 
evolving in $(x,t)$ plane. We emphasis that this breather solution has periodic singularities in $(x,t)$ plane for condition $\phi(x,t)=0$. Hence this singular solution has not a direct application for physical problems. However, such solutions are important for the general theory of nonlinear systems \cite{Mat}.

In conclusion, we have presented several types of spatially localized and
periodic wave solutions for the mKdV equation describing propagation of
nonlinear waves in many physics areas when there is polarity symmetry.
A novel class of exact soliton solutions given by Eq. (\ref{8}) is firstly obtained for the model. Thise solutions describe the propagation of a pair interacting solitons with
opposite polarity, which interact locally and behave like independent
solitons far from the meeting point. We have also found a number of exact 
interacting periodic wave solutions for the mKdV  equation.
The results showed that the formation of those closed form solutions is
determined by the sign of cubic nonlinearity parameter solely. Undoubtedly,
these new solutions will be useful for recognizing physical phenomena and
dynamical processes in various physical systems where the mKdV equation can
provide a realistically accurate description of the waves.


\begin{thebibliography}{99}
\bibitem{R1} H. Ono, J. Phys. Soc. of Japan \textbf{61}, 4336 (1992).

\bibitem{R2} E. A. Ralph and L. Pratt, J. Nonlinear Science \textbf{4}, 355
(1994).

\bibitem{R3} T. S. Komatsu and Shin-ichi Sasa, Phys. Rev. E \textbf{52},
5574 (1995).

\bibitem{R4} H. X. Ge, S. Q. Dai, Y. Xue and L. Y. Dong, Phys. Rev. E 
\textbf{71}, 066119 (2005).

\bibitem{R5} Z.-P. Li and Y.-C. Liu, European Phys. J. B - Condensed Matter
and Complex Systems \textbf{53}, 367 (2006).

\bibitem{R6} A. H. Khater, O. H. El-Kalaawy and D. K. Callebaut, Physica
Scripta \textbf{58}, 545 (1998).

\bibitem{R7} K. E. Lonngren, Optical and Quantum Electronics \textbf{30},
615 (1998).

\bibitem{R8} V. Ziegler, J. Dinkel, C. Setzer and K. E. Lonngren, Chaos,
Solitons \& Fractals \textbf{12}, 1719 (2001).

\bibitem{R9} R. Grimshaw, E. Pelinovsky, T. Talipova, Nonlinear Processes in
Geophysics. \textbf{4}, 237 (1997).

\bibitem{R10} H. Leblond and D. Mihalache, Phys. Rev. A \textbf{79}, 063835
(2009).

\bibitem{R11} H. Leblond and D. Mihalache, J. Phys. A: Math. Theor. \textbf{%
43}, 375205 (2010).

\bibitem{R12} H. Triki, H. Leblond, D. Mihalache, Opt. Commun. \textbf{285, }%
3179 (2012).

\bibitem{R13} H. Leblond, H. Triki, and D. Mihalache, Rom. Rep. Phys. 
\textbf{65}, 925 (2013)

\bibitem{R14} H. Leblond, Ph. Grelu and D. Mihalache, Phys. Rev. A \textbf{90%
}, 053816 (2014).

\bibitem{Pel} D. Pelinovsky and R. Grimshaw, Phys. Lett. A \textbf{229}, 165
(1997).

\bibitem{Wadati} M. Wadati, J. Phys. Soc. Japan \textbf{32}, 1681(1972).

\bibitem{Kevrekidis} P. G. Kevrekidis, A. Khare, A. Saxena and G. Herring,
J. Phys. A: Math. Gen. \textbf{37}, 10959 (2004).

\bibitem{Lamb} J. L. Lamb, \textit{Elements of Soliton Theory} (John Wiley
\& Sons, New York, 1980).

\bibitem{G} R. Grimshaw, E. Pelinovsky, T. Talipova, M. Ruderman, and R. Erd%
\'{e}lyi, Stud. Appl. Math. \textbf{114}, 189 (2005).

\bibitem{An} A. Ankiewicz, J. M. Soto-Crespo, and N. Akhmediev, Phys. Rev. E 
\textbf{81}, 046602 (2010).

\bibitem{Zhang} G. Zhang and Z. Yan, Physica D \textbf{410}, 132521 (2020).

\bibitem{Miur} R. M. Miura, C. S. Gardner, M. D. Kruskal, J. Math. Phys. 
\textbf{9}, 1204 (1968).

\bibitem{Pego} R. L. Pego and M.I. Weinstein, Philos. Trans. R. Sot. A 
\textbf{340}, 47 (1992).

\bibitem{Mat} V. B. Matveev, M. A. Salle, \textit{Darboux transformations
and solitons}. Springer-Verlag, 1991.

\bibitem{Slu} A. V. Slunyaev, J. Exp. Theor. Phys. \textbf{92}, 529 (2001).

\bibitem{Slunyaev} A. V. Slunyaev and E. N. Pelinovsky, Phys. Rev. Lett. 
\textbf{117}, 214501 (2016).
\end{thebibliography}
\end{document}